\documentclass[twocolumn,showpacs,reprint,preprintnumbers,prl,aps,10pt]{revtex4-1}
\usepackage{amsmath,amsthm,amssymb,amscd,epsfig}
\usepackage{graphicx}
\usepackage{psfrag}
\usepackage{amsmath}
\usepackage{color}

\newcommand {\rd}{\mathrm{d}}

\newcommand {\eff}{\mathrm{eff}}

\begin{document}

\title{Dynamical mechanisms leading to equilibration in two-component
gases}
\author{Stephan~De~Bi\`{e}vre}
\email{stephan.de-bievre@univ-lille1.fr}
\affiliation{Laboratoire Paul Painlev\'e, CNRS, et UFR de Math\'ematiques,
Universit\'e Lille 1; \'Equipe-Projet Mephysto, INRIA Lille-Nord Europe, France}
\author{Carlos~Mej\'{\i}a-Monasterio}
\email{carlos.mejia@upm.es}
\affiliation{Laboratory of Physical Properties, Technical University
of Madrid, Av. Complutense s/n 28040 Madrid, Spain}
\author{Paul~E.~Parris}
\email{parris@mst.edu}
\affiliation{Department of Physics, Missouri University of Science \& Technology, Rolla,
MO 65409, USA}

\begin{abstract}
Demonstrating how microscopic dynamics cause large systems to approach thermal equilibrium remains an elusive, longstanding, and actively-pursued goal of statistical mechanics. We identify here a dynamical mechanism for thermalization in a general class of two-component dynamical Lorentz gases, and prove that each component, even when maintained in a non-equilibrium state itself, can drive the other to a thermal state with a well-defined effective temperature.
\end{abstract}

\maketitle

%05.70.-a: Thermodynamics
%02.50.Ey: Stochastic processes

That   isolated  systems   with   many  degrees   of  freedom   evolve
asymptotically in  time towards thermal equilibrium lies  at the heart
of  classical thermodynamics. Statistical  mechanics teaches  that for
systems described by  a Hamiltonian $H$, the thermal  states are those
described  by the  canonical Boltzmann  relation $\rho  =$ $Z^{-1}\exp
(-\beta H)$.  This follows from original arguments of Maxwell, marginal
distributions  that  arise  from  microcanonical  ensembles,  and  the
properties of the maximum   entropy   states    to   which
systems   thermodynamically tend~\cite{Khinchin}.

While  such statistical arguments  identify  the thermal  state,  they provide  no
insight into  the problem of  how the microscopic dynamics  of diverse
large systems each lead  towards equilibrium from an arbitrary initial
state.

Considerable progress  has been made in  understanding \emph{return to
  equilibrium}~\cite{return}  in which  a  small system  coupled to  a
large  thermal  reservoir  thermalizes   to  the  temperature  of  the
latter. The more general problem of \emph{approach to equilibrium}, in
which  many mutually interacting  elements of  an isolated  system are
initially out of  equilibrium, is less understood~\cite{approach}.  In
spite   of   recent   progress,   including  identification   of   the
thermalization mechanism  in the Fermi-Pasta-Ulam problem~\cite{ovpl},
there is continuing discussion and even some controversy regarding the
role  played by  various properties  of the  dynamics, such  as chaos,
mixing, and resonances \cite{casas2003, polkovnikov2011}, as well
as on the emergence of non-equilibrium
effective temperatures~\cite{ritort}.

As an important step in this longstanding problem,
we identify here a dynamical mechanism for thermalization
in a general class of two component systems~\cite{llm,llm2,em,cmp,dmp2,dp,spd,dpl,adlp},
in   which
non-interacting point particles move  freely through a spatially fixed
array of  isolated dynamical ``scatterer'' particles, each of which has a few
rotationally-invariant  degrees   of  freedom  with   which
the itinerant point particles
interact and exchange energy when they are within a fixed range.

In particular, we prove here that, starting from an arbitrary initial state, an
ensemble  of probe
subsystems (either the itinerant point particles or the ``scatterer'' particles)
when subjected to repeated
weak interactions with members of the complementary component gas,
acting as a homogeneous
and stationary energy reservoir, approaches  a Boltzmann state at a well-defined temperature,
provided  merely that  the  reservoir  is
stationary--\emph{it need  not be  in thermal equilibrium.}

While providing an understanding of the approach to equilibrium of the entire
system as a whole, the result also provides insight into and allows a justification of
the concept of non-equilibrium effective temperatures.

In systems of this type  studied previously, the scatterers were taken
to  be disks~\cite{klages, llm,llm2,em,cmp}  or  needles in  2D~\cite{dmp2}
that     rotate     about      fixed     centers,     and     harmonic
oscillators~\cite{dp,spd,dpl,adlp}.  In  each  of these  systems,  any
particle or  scatterer in the gas follows  a non-interacting Newtonian
evolution   during  time  intervals   separating  the   collisions  it
experiences. In each collision an energy conserving interaction occurs
between the  particle and scatterer  involved.  Such systems, in which
the scatterers themselves possess an internal dynamics, have 
been referred to as dynamical Lorentz gases \cite{dp}, since  
they provide a natural generalization of the two-component gases 
originally introduced by Lorentz \cite{lorentzgas}, and subsequently 
studied by many others 
\cite{d}, that feature \emph{inert} (i.e., non-dynamical) scatterers, and which 
for over a century have been an essential tool for understanding diffusion, 
related equilibrium and non-equilibrium statistical properties, and 
the so-called Boltzmann-Grad limit.

We focus  here on thermalization in a specific two-dimensional system  of this
type, introduced in  \cite{llm} as the rotating Lorentz  gas model (RLG),
in which the scatterers are rotating disks.   The full many particle
RLG exhibits realistic equilibrium and
non-equilibrium  behaviour~\cite{llm2},  and has  been  used to  study
thermal rectification~\cite{em} and thermoelectricity~\cite{cmp}.  Its
time-reversible dynamics preserves phase space volume.

To understand  equilibration of such  RLGs as a whole,  we investigate
here  the   effect  of  repeated  interactions  on   the  phase  space
distribution of  a \emph{single member}  of each of the  two
components making up the RLG. {In the context of
non-equilibrium temperatures, one might think of
an ensemble of such single members as a ``thermometer'' locally probing the phase space
distribution of the other component.}
\begin{figure}
\includegraphics[width =3in]{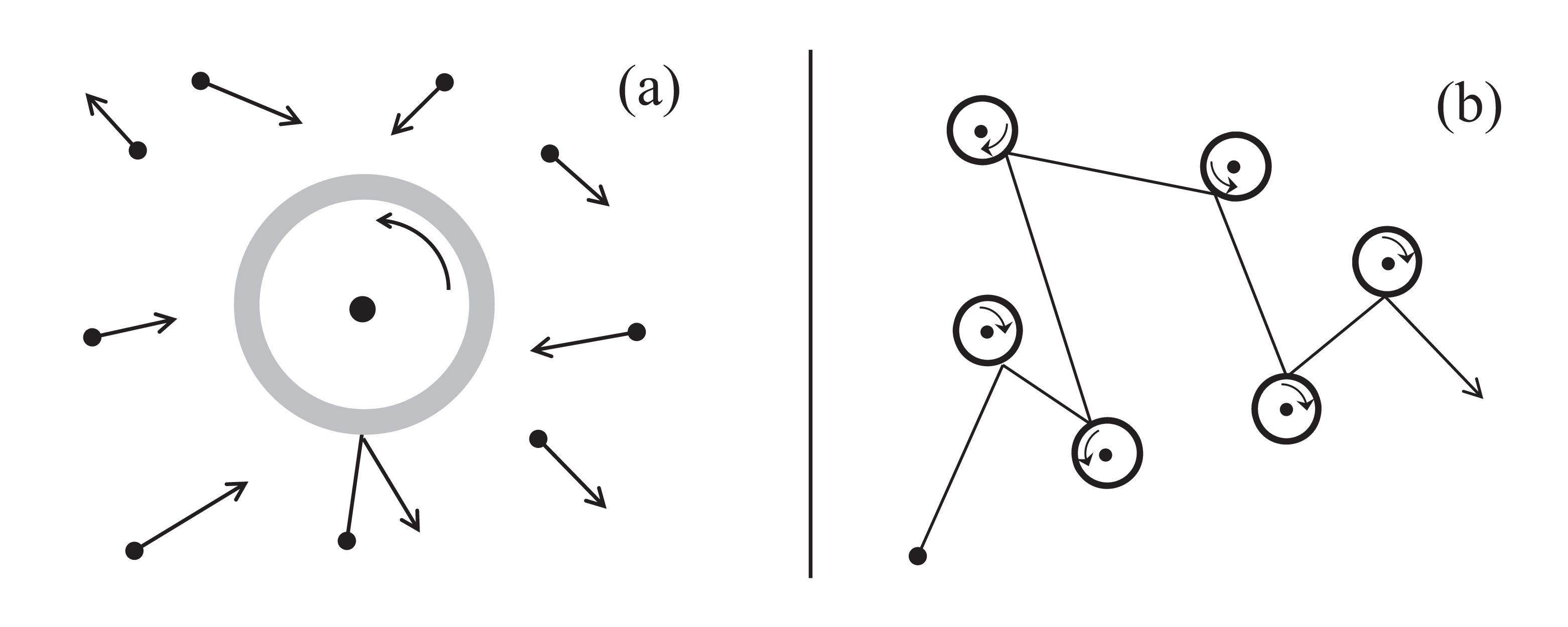}
\caption{(a) A single rotating disk in a gas of point particles, and
(b) a single point particle passing through a spatially fixed
array of rotating disks.
\label{fig:1}}
\end{figure}

Thus, we first consider a single hollow disk that rotates
freely  about its  center immersed in  a  gas of  (and subjected  to
repeated     impacts    by)     non-interacting     point    particles
(Fig.~\ref{fig:1}a),  with the  particles of  the gas  reservoir being
``probed'' drawn independently from  a  stationary, but not  necessarily
thermal distribution.

This  obviously  ignores   recollisions,  which  would modify  the  momentum
distribution  of  gas particles in the full RLG as they  repeatedly
encounter  different scatterers.  In the second part  of our analysis, therefore, we follow
a   single  particle   as   it  passes   through   (i.e.,  probes)   a
spatially-fixed gas  of rotating disks  (Fig.~\ref{fig:1}b), where now
it is the disks of  the scatterer reservoir that are drawn independently from an
arbitrary stationary distribution.

{As we prove, in either of these situations, the probe  dynamics reduce to  a Markov
chain that leads in the limit of small average energy exchange to
an approach of the corresponding probe energy distribution
to a Boltzmann state with a
well-defined effective temperature.}

\begin{figure}[tbp]
\includegraphics[width=2.35in]{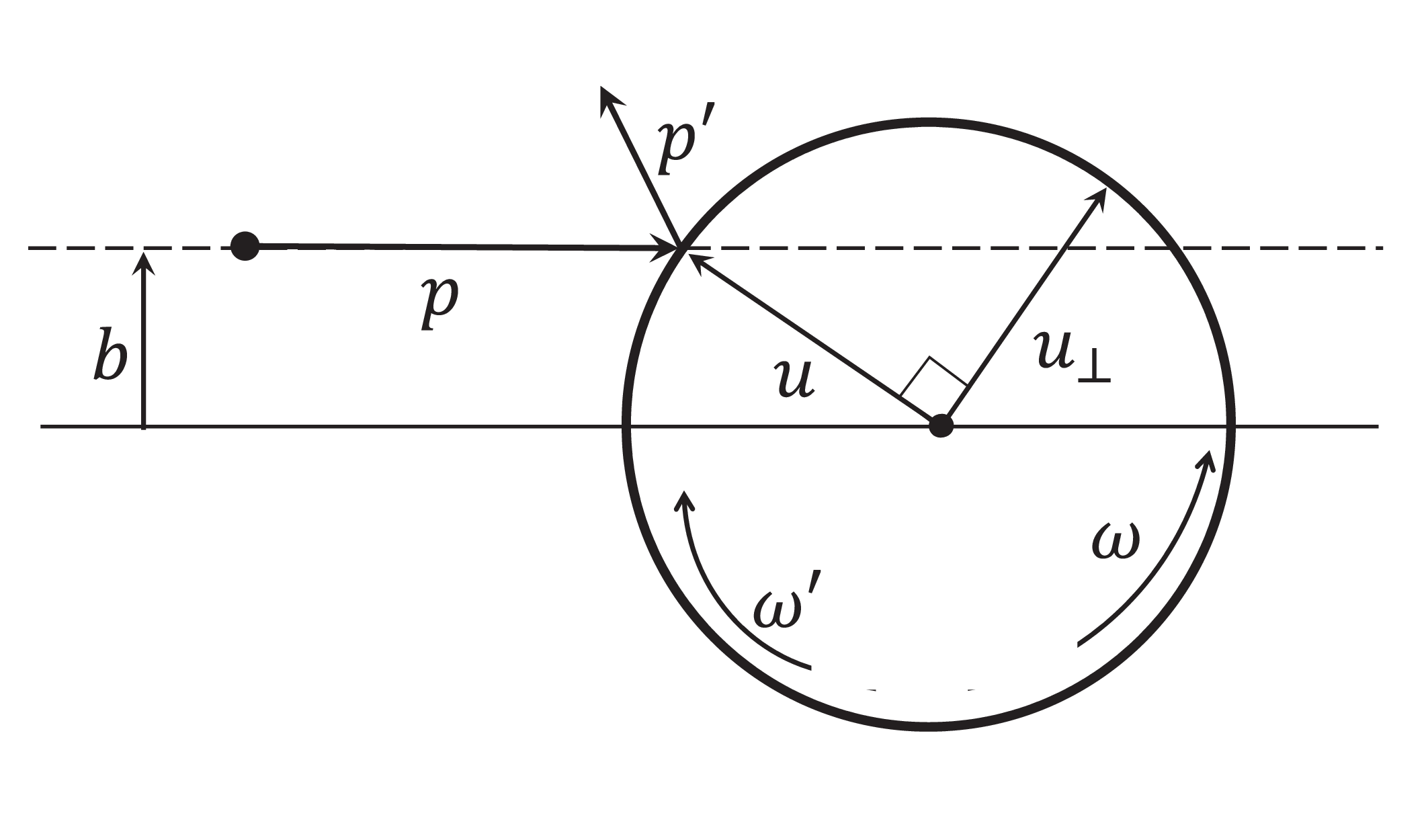}
\caption{Collision geometry for a point particle of unit mass
and initial momentum $p$ impinging on a freely rotating disk of
mass $\mu $ and initial angular velocity $\omega$.
\label{fig:collision}}
\end{figure}

In either case, each    step    of     Markov chain involves     a    collision
(Fig.~\ref{fig:collision})  between  a  particle  of  mass  $m=1$  and
initial momentum $p$ impinging with impact parameter $b$ upon a hollow
disk  of unit  radius and  mass  $\mu=M/m$, rotating  about its  fixed
center with initial angular  velocity $\omega$.  During the collision,
particle and disk obtain new momentum and angular velocity~\cite{llm}
\begin{eqnarray}
p^{\prime}&=&p-2(p\cdot u)u-\frac{2\mu }{1+\mu }
\left( p_{t}-\omega\right) u_{\perp },  \label{eq:mexicanrule1} \\
\omega^{\prime}&=&\omega+\frac{2}{1+\mu}(p_{t}-\omega).
\label{eq:mexicanrule2}
\end{eqnarray}
Here, $\mu$  represents both the  mass of the  disk and its  moment of
inertia, the  unit vector $u$ links  the disk center  to the collision
point, $u_{\perp }$ is obtained  by rotating $u$ through $\pi /2$, and
$p_{t}=p\cdot u_{\perp}=b  \|p\| $.   Under these rules,  particle and
disk exchange energy and angular momentum, conserving both.

\paragraph*{One rotating disk in a gas of particles~~--}
Consider  a  single disk  (the ``thermometer'') with  initial  angular velocity  $\omega_0$,
subject  to  impacts   by  point  particles  (Fig.~\ref{fig:1}a)  with
i.i.d.~momenta $p_n$ drawn  from a rotationally invariant distribution
$\rho_{\eff}(|p|)$,  with i.i.d.~impact  parameters  $b_n$ uniform  in
$[-1,1]$, at  a Poissonian sequence  $t_n$ of impact times.  Note, the
marginal   distribution  $\rho_{\eff}(|p|)\propto\rho_{0}(|p|)|p|$  of
particles striking the disk, related  to the ``effusive'' flux $j(p) =
\rho_{0}(|p|)p$ of particles  on the disk, is different  from the bulk
gas distribution $\rho_{0}(|p|)$.

Under  these conditions $\langle  p_t\rangle=0.$ After  $n$ collisions
~\eqref{eq:mexicanrule2} gives
\begin{equation}
\omega_n=\gamma^n\omega_0+\eta\sum_{k=0}^{n-1}\gamma^k p_{t, n-1-k},
\label{eq:omega_n}
\end{equation}
where   $\eta=2/(1+\mu   )$   and   $\gamma=1-\eta\in   (-1,1)$.    As
$n\to\infty$, the first term in \eqref{eq:omega_n} vanishes, leaving a
convergent sum
\begin{equation}
\omega = \lim_{n\to+\infty}\eta\sum_{k=0}^{n-1}\gamma^k p_{t, n-1-k}
\label{eq:omega_infinity}
\end{equation}
of   \emph{geometrically  weighted}  i.i.d.~random   variables  $p_t$,
ensuring that  the angular  velocity distribution approaches  a stationary
limit $\rho (\omega )$.

When  the   particle  gas  is   in  thermal  equilibrium,   the  $p_t$
in~\eqref{eq:omega_infinity} are Gaussian.   Hence, independent of the
value $\mu$, the distribution $\rho  (\omega )$ is thermal and has the
same  temperature as the  bath (see  below).  Thus, in contact with
thermal particles, a probe disk approaches a common
thermal equilibrium with the particle bath.

\begin{figure}[tbp]
\includegraphics[width=0.475\textwidth]{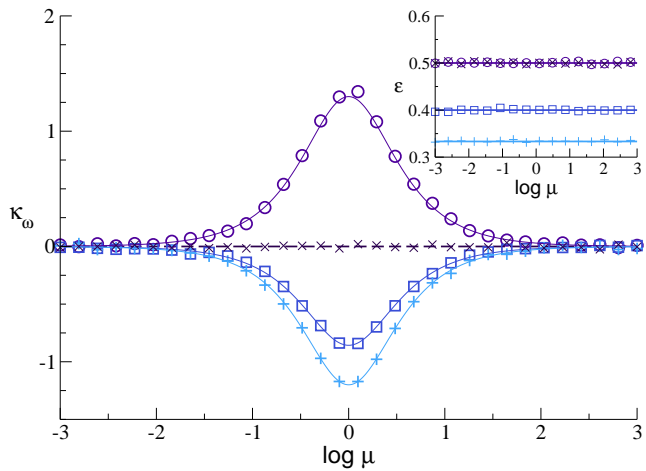}
\caption{Limiting value of $\kappa_\omega$ of
a disk of mass $\mu$  subjected  to repeated
impacts by  particles   drawn   from  thermal ($\times$),
uniform ($\Box$), and microcanonical ($+$) distributions,
and  a superposition  of  two
thermal  distributions ({\large $\circ$}) with
temperatures
in  ratio $T_{1}/T_{2}  = 5$.  Curves  are  theoretical  predictions,
symbols are simulation results.
Inset: Limiting value of kinetic energy ratio
$ \varepsilon  = \varepsilon_{\rm{d}} / \varepsilon_{\rm{p}} $
vs  $\mu$. Theoretical
predictions are $\varepsilon=1/2, \varepsilon=2/5$,
and $\varepsilon=1/3$.
\label{fig:disk_kurtosis}}
\end{figure}

Less    obvious   and more interesting is    when   the    particle   reservoir    is   not
thermal~\cite{dmp2}. Here,  we characterize the  limiting distribution
by  its low order  moments. Assuming  $\langle\omega_0\rangle=0$, then
$\langle \omega_n\rangle=0$, and
\begin{equation}
\langle\omega_n^2\rangle =\gamma^{2n}\langle \omega_0^2\rangle +\eta^2\langle
p_t^2\rangle \left(\frac{1- \gamma^{2n}}{1-\gamma^2}\right).
\end{equation}
As $n\to\infty$, the average disk energy then approaches
\begin{equation}
\varepsilon_{\rm{d}} = \frac{\mu}{2}\langle \omega ^2\rangle
=\frac{1}{2}\langle p_t^2\rangle
= \frac{1}{6}\langle p^2\rangle_\eff.
\label{eq:mean_energy}
\end{equation}
On  the  right,  the  average  $\langle b^2\rangle  =  1/3$  has  been
performed,  and the  remaining average  is over  $\rho_\eff(|p|)$.  To
relate this  to the particle  energy $\varepsilon_{\rm{p}}=\langle p^2
\rangle_{0}/2$ [where $\langle  \ldots \rangle_{0}$ indicates averages
over  $\rho_{0}(|p|)$],  we  introduce  an energy  partitioning  ratio
$\varepsilon   =   \varepsilon_{\rm{d}}/\varepsilon_{\rm{p}}$,   which
equals $1/2$   {in equilibrium, satisfying the equipartition theorem}.
For a  disk immersed  in a  non-equilibrium
particle  gas,  however, $\varepsilon  =  \langle p^2  \rangle_{\eff}/
\langle 3p^{2} \rangle_{0}$ approaches a non-universal value
that depends on  the particle distribution, but is  independent of the
mass of  the disk  (see Fig.~\ref{fig:disk_kurtosis}).  For  a thermal
particle bath, $\varepsilon  = 1/2$ and~\eqref{eq:mean_energy} reduces
to the equilibrium result $\mu \langle \omega^{2}\rangle /2 = k_{B}T /
2$.

To study the \emph{shape} of $\rho (\omega)$ we compute the asymptotic
value of the \emph{excess disk kurtosis}
\begin{equation}
\kappa_{\omega} =
\beta_2-3=\eta \frac{1+\gamma}{1+\gamma^2}\left(\frac{\langle p_t^4\rangle}
{\langle p_t^2\rangle^2}-3\right),
\label{eq:kurtosis}
\end{equation}
in
which  $\beta_2=\langle \omega ^4\rangle/\langle  \omega ^2\rangle^2$, and which vanishes when the limiting  disk distribution is thermal.
This is the case for a thermal particle bath, when the excess
particle   kurtosis   $\kappa_{\rm{p}}=\left(\langle  p_t^4\rangle   /
  \langle p_t^2\rangle^2-3\right)$ in \eqref{eq:kurtosis} vanishes.

But $\rho (\omega)$ \emph{also} reduces to a thermal distribution when
the particle reservoir is \emph{not} thermal, whenever the coupling is
sufficiently weak. This  occurs when the disk is  very heavy ($\mu \gg
1$, $\eta\to 0$, $\gamma\to 1$), and very light ($\mu \ll 1$, $\eta\to
2$, $\gamma\to -1$).  In these two limits~\eqref{eq:kurtosis} vanishes
and $\rho (\omega)$ becomes Gaussian~\cite{z}.  In weak coupling, this
thermalization is  universal, i.e., \emph{independent} of  the form of
$\rho_{\eff}(|p|)$.

Thus, when placed weakly in contact with a stationary, non-equilibrium
particle gas,  the probe disk ``thermalizes to''  (or ``measures'') an
apparent  temperature such that  $k_{B}T/2 =  \varepsilon \left\langle
  p^{2}  \right\rangle_{0}/2,$   where  $\varepsilon  =   \langle  p^2
\rangle_{\eff}/  \langle  3p^2 \rangle_{0}$  depends  on the  particle
distribution and is generally  not equal to $1/2$. This thermalization
for    large     and    small    $\mu$    is     clearly    seen    in
Fig.~\ref{fig:disk_kurtosis},  which displays a  numerical computation
of  the  values of  $\varepsilon$  and  $\kappa_{\omega}$, for  $10^5$
disks,  each  subjected  to  $10^4$ repeated  collisions  for  various
particle distributions.

From  Fig.~\ref{fig:disk_kurtosis} and  Eq.~\eqref{eq:kurtosis}  it is
clear  that  for  intermediate  and  strong  coupling,  with  particle
reservoirs  for   which  $\kappa_{\rm{p}}  \neq  0$,   the  disk  does
\emph{not}  approach  a   thermal  state.   Instead,  repeated  strong
interactions drive it to a non-thermal state with $\kappa_{\omega}\neq
0$.

\paragraph*{One particle in a gas of disks~~--}
We now  consider a  single particle that  collides with a  sequence of
rotating  disks  (Fig.~\ref{fig:1}b)   whose  angular  velocities  are
i.i.d.~variables   drawn   from   a  stationary   distribution   $\rho
(|\omega|)$,  again  ignoring  recollisions.   Denote by  $p_{s}$  the
particle's momentum before collision $s$, when it impinges with impact
parameter $b_s$ on a disk with angular velocity $\omega_s$.  According
to \eqref{eq:mexicanrule1}, after this collision
\begin{equation}
p_{s+1}=p_{s}-2(p_{s}\cdot u_{s})u_{s}-\frac{2\mu }{1+\mu }\left(
p_{s,t}-\omega _{s}\right) u_{s,\perp }.
\label{eq:mexicanrule}
\end{equation}%
Collisions occur at times $t_{s+1}=t_{s}+\ell /p_{s}$, where $\ell$ is
the collision mean free path.  We assume $\langle p_0 \rangle = 0$.

The Markov chain \eqref{eq:mexicanrule} for $p$ is less tractable than
for a single  disk in a gas of particles.   In weak coupling, however,
analysis of the Markov chain to determine the limiting distribution is
straightforward.
\begin{figure}[tbp]
\includegraphics[width=0.475\textwidth]{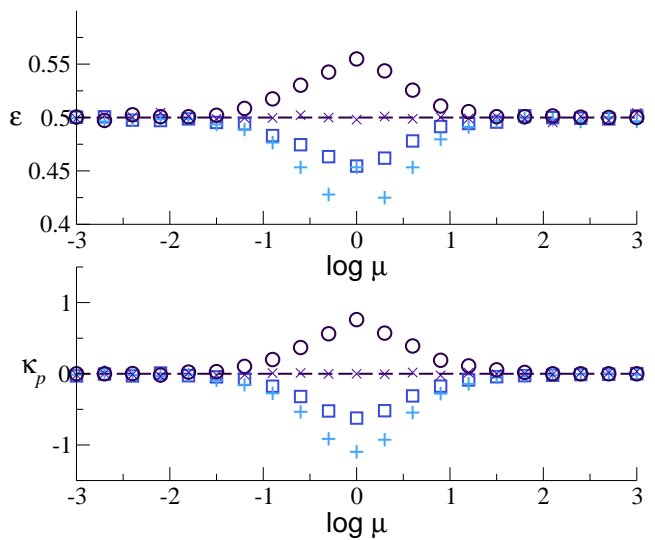}
\caption{Limiting values of the kinetic energy ratio
$ \varepsilon  = \varepsilon_{\rm{d}} /
\varepsilon_{\rm{p}} $
and excess kurtosis $\kappa_{\rm{p}}$ for a particle
colliding with rotating disks
of mass $\mu$ and
angular velocities drawn from thermal ($\times$), uniform
($\Box$), and microcanonical ($+$) distributions,
and a superposition  of  two thermal
distributions ({\large $\circ$}) with
$T_{1}/T_{2}  = 5$.
\label{fig:particle_kurtosis}}
\end{figure}

\begin{figure}[tbp]
\includegraphics[width=0.475\textwidth]{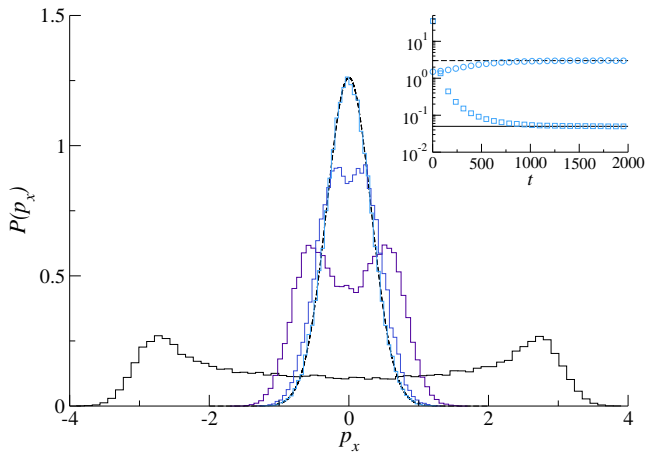}
\caption{Distribution function for velocity component
  $p_x$ for a particle undergoing repeated collisions
  with rotating disks of mass
$\mu=0.005$ in a microcanonical ensemble with $|\omega |= \sqrt{20}$,
recorded (from broad to narrow) at times $t=60, 300, 600,
2000$. Dashed curve corresponds to the limiting thermal
distribution. Inset: convergence of the
particle kurtosis $\beta_2^{(p)}$  ({\large $\circ$}) to its limiting
value $3$ (dashed line), and
kinetic energy ratio $ \varepsilon  = \varepsilon_{\rm{d}} /
\varepsilon_{\rm{p}} $ ($\Box$) to its limiting value $1/2$ (solid line).
\label{fig:particle_convergence}}
\end{figure}

Figures~\ref{fig:particle_kurtosis} and~\ref{fig:particle_convergence}
display simulation  results in which  $10^5$ particles each  undergo a
sequence of  collisions with rotating disks  initialized as described.
In  Fig.~\ref{fig:particle_kurtosis},   values  of  $\varepsilon$  and
$\kappa_{\rm{p}}$ for the  limiting particle distribution, recorded at
long   fixed  simulation   time,  are   plotted  for   different  disk
distributions $\rho(|\omega |)$.

For thermal disks,  the particle equilibrates to a  thermal state with
the  same temperature.  Moreover, when  $\mu$  is very  large or  very
small, independent of $\rho(\omega )$,  the particle is also driven to
a thermal state (see, e.g., Fig.~\ref{fig:particle_convergence}), with
a    vanishing   $\kappa_{\rm{p}}$.     Thus,    we   again    observe
``thermalization''  of   a  small  system  in  weak   contact  with  a
non-thermal reservoir.   However, unlike what  happens to a disk  in a
gas  of  particles, the  limiting  particle temperature  \emph{always}
obeys  equipartition,  since  $\varepsilon  \rightarrow 1/2$  in  this
regime.
For  intermediate  $\mu$,   provided  $\kappa_{\omega}  \neq  0$,  the
particle does not generally thermalize.

To analytically  demonstrate the thermalization observed  at large and
small  $\mu$, we  express using  \eqref{eq:mexicanrule} the  change in
particle energy
\begin{eqnarray}  \label{eq:col-rule}
E_{s+1}- E_s &=&  - \frac{4\mu}{(1+\mu)^2} E_s b_s^2 + \frac{2\mu}{(1+\mu)^2}
\mu\omega_s^2  \notag \\
&\ &\quad + \quad 2\sqrt{\mu} \frac{1-\mu}{(1+\mu)^2}\sqrt{2 E_s} \sqrt{\mu}%
\omega_s b_s
\end{eqnarray}
during collision $s$, in terms  of its energy immediately before.  The
coefficients       multiplying      the       dynamical      variables
in~\eqref{eq:col-rule}  set the  scale for  the energy  change  in any
collision, and  are small  for $\mu \ll  1$ and  $\mu \gg 1$.   In the
latter case the  equation for large $\mu$ follows  from that for small
$\mu$ by replacing $\mu$ with $\mu^{-1}$.

It suffices to study  this weak-coupling, small-step limit for $\mu\ll
1$.  In this limit,~\eqref{eq:col-rule} becomes
\begin{equation}  \label{eq:wc-limit}
\Delta E_{s} = -4\mu E_s b_s^2 + 4\mu \varepsilon_{\rm{d}} k_s^2
+ 2\sqrt{4\mu}\sqrt{\varepsilon_{\rm{d}} E_s} k_s b_s,
\end{equation}
where
\begin{equation}
\varepsilon_{\rm{d}} = \frac{1}{2}\mu \langle \omega_{s}^{2}\rangle ,\quad
k_{s} = \sqrt{\frac{2\mu}{\varepsilon_{\rm{d}}}} \omega_{s} , \quad
\langle k_{s}^{2} \rangle = 1.
\label{eq:ks}
\end{equation}
Averages here are over $\rho(|\omega  |)$.  The first two terms on the
right  hand  side  of   \eqref{eq:wc-limit}  constitute  a  source  of
``dynamical    friction''  \cite{Cha43a,Cha43b} that   
counterbalances    the   stochastic
acceleration caused by the fluctuating last term~\cite{dp,adlp, Cha43a,Cha43b}.
This competition thus leads to a kind of fluctuation-dissipation 
like mechanism that naturally emerges from the deterministic dynamics.

Inspection  of~\eqref{eq:wc-limit} suggests  $\xi =  4\mu$ as  a small
parameter.  Introducing (scaled) continuous collision number $\sigma =
\xi  s$, Eq.~\eqref{eq:wc-limit}  can be  described by  the stochastic
differential equation
\begin{equation*}  \label{eq:SDE}
dE(\sigma ) = - \alpha(E)d\sigma + \lambda(E)dw(\sigma )
\end{equation*}
in which
\begin{equation*}  \label{eq:drift}
\alpha(E) = \frac{1}{3}E(\sigma) - \varepsilon_{\rm{d}}\ ,
\quad \lambda(E) = \frac{2}{\sqrt{3}}\sqrt{\varepsilon_{\rm{d}}E(\sigma)}  \ .
\end{equation*}
Here  $dw$  is  white  noise,  $\langle dw\rangle  =0$,  and  $\langle
dw(\sigma)dw(\sigma ^\prime) \rangle = \delta(\sigma-\sigma ^\prime)$.
Clearly,  the results will  depend on  only the  first two  moments of
$\rho(|\omega |)$.

From   the  corresponding  Fokker-Planck   equation,  one   finds  the
stationary distribution
\begin{eqnarray*}
f_\infty(E)&=&\tilde{Z}^{-1}\lambda^{-2}(E)
\exp [~-\int 2\alpha(E) \lambda^{-2}(E) \rd E~] \\
&=&\frac1{Z'}E^{1/2}\exp(-\frac{E}{2\varepsilon_{\rm{d}}})
\end{eqnarray*}
for $E(\sigma)$ at large  $\sigma$.  Physically, $f(E,\sigma)\rd E$ is
the fraction of particles after collision $\sigma$ with energy between
$E$ and $E+\rd E$.  Such particles stay in that state for time $\Delta
t=  \ell  /\sqrt{2E}$.   Thus,  at  long times,  the  distribution  of
particles with energy $E$ becomes thermal
\begin{equation}\label{eq:boltzmann!}
f(E)=\frac1Z \exp(-\frac{E}{2\varepsilon_{\rm{d}}})
\end{equation}
as  in a  2D non-interacting  particle gas  with  apparent temperature
given by  $k_{B}T=2\varepsilon_{\rm{d}}$. Thus, rotating  disks not in
equilibrium  drive the  particle to  a thermal  state  compatible with
equipartition.

\paragraph*{On the approach to equilibrium~~--}
In weak  coupling, as we have  shown, both components of  the
full {RLG} are  driven    to     a    Gaussian     thermal     state.
From
\eqref{eq:mexicanrule}, and \eqref{eq:wc-limit}, 
it is  clear that equilibration is not a  
simple consequence  of the central limit theorem.  Indeed,  the momentum 
or energy increments that particles experience are not i.i.d. 
random variables; they are steps 
in an associated Markov chain, with the size and variance of each being a 
function of the dynamical variables at each step.  
Thus, the 
limiting exponential distribution for the particle energy in this case can be 
understood as arising from a competition between impulsive fluctuating 
forces of zero mean, which tend to heat the particle, and
a dynamical friction that leads higher energy particles to lose energy 
to the disks. Thus, a fluctuation-dissipation 
mechanism emerges naturally from the deterministic dynamics associated 
with the interaction between the particles and the rotating disks. 

%{\red
%In that work, the dissipative part of the particle evolution emerged from
%a kind of \emph{back reaction} in which the motion of the oscillators was 
%altered during the (finite) interaction time by the presence of the particles 
%in a way that further altered the motion of the particles themselves. 
%FIX FROM HERE In the present model the same thing happens. if $\mu=+\infty$, stochastic acceleration. it's the same! back reaction is here too.
% }

The fact that in the present case both components
of the full {RLG} equilibrate in the presence of a non-thermal
bath allows us to identify
the mechanism of approach to
thermal equilibrium of the {RLG} as a whole.

   At low particle density,
e.g., each  particle will  scatter off many  disks before any  disk is
likely to  have interacted  with more than  a few particles.   In weak
coupling, therefore, the particle gas will equilibrate well before the
gas of disks, to a  thermal distribution in which equipartition of the
particle and disk energies is obtained.  Each disk will then be in contact
with a thermal distribution of particles, and will only need therefore
to  undergo  a  \emph{return}  to  the  appropriate  limiting  thermal
distribution.   Beyond  weak  coupling,  the  situation  becomes  more
complicated \cite{dmp2}.

In summary, we have identified a thermalization mechanism in 
 a rotating disk, weak-coupling version of a two-component dynamical 
 Lorentz gas. We expect the basic underlying dynamical friction mechanism to be effective
 more generally in systems in which the individual components undergo repeated scattering events (See, for example 
 Ref. \onlinecite{dp})
 
 %,  a similar fluctuation-dissipation mechanism due to dynamical friction
% was shown to lead to equilibration of a gas of mobile particles moving through a gas of 
%fixed harmonic oscillators.}

% that shows features in common with behavior previously identified in 
% weak coupling for a similar model in which the scatterers are 
% harmonic oscillators. 
% 
% 
% 
% Clearly, in each case the specific results 
% obtained depend in detail on the particular dynamics of the individual
% models. The common features and the dynamical behavior that they share
% nonetheless suggests, as one intuitively expects, that 
% similar principles ought to apply more generally.}
%

\begin{acknowledgments}
Part of this work was performed while
C.M.-M. and P.E.P. visited the Universit\'e Lille 1 and the Labex CEMPI
(ANR-11-LABX-0007-01).  They   thank  those  institutions   for  their
hospitality. C.M.-M. acknowledges partial financial support from the Spanish MICINN
  grant MTM2012-39101-C02-01 and from ONRG Grant N62909-15-1-C076.
\end{acknowledgments}

\end{document}